\documentclass[%
 aps,
 amsmath,amssymb,
 reprint,%
superscriptaddress
]{revtex4-1}

\usepackage{graphicx}% Include figure files
\usepackage{dcolumn}% Align table columns on decimal point
\usepackage{bm}% bold math
\usepackage[utf8]{inputenc}
\usepackage[T1]{fontenc}
\usepackage{mathptmx}
\usepackage{xcolor}
\usepackage{ulem}
\begin{document}

\preprint{AIP/123-QED}

%\title{Tunable polarization of broadband spintronic terahertz emitter based on magnetization textures}
\title{Controlling polarization of spintronic THz emitter by remanent magnetization texture}
% Force line breaks with \\

\author{Weipeng Wu}
\affiliation{Department of Physics and Astronomy, University of Delaware, Newark, Delaware 19716, USA}%
\author{Sergi Lendinez}
\affiliation{Department of Physics and Astronomy, University of Delaware, Newark, Delaware 19716, USA}%
\author{Mojtaba Taghipour Kaffash}
\affiliation{Department of Physics and Astronomy, University of Delaware, Newark, Delaware 19716, USA}%
\author{Richard D. Schaller }
\affiliation{Center for Nanoscale Materials, Argonne National Laboratory, Argonne, Illinois 60439, USA}%
\affiliation{Department of Chemistry, Northwestern University, Evanston, Illinois 60208, USA
}%
\author{Haidan Wen}
\affiliation{Advanced Photon Source, Argonne National Laboratory, Argonne, Illinois 60439, USA}%
\author{M. Benjamin Jungfleisch}
\email[]{mbj@udel.edu}
\affiliation{Department of Physics and Astronomy, University of Delaware, Newark, Delaware 19716, USA}%

\date{\today}% It is always \today, today,

\begin{abstract}
{Terahertz (THz) sciences and technologies have contributed to a rapid development of a wide range of applications and expanded the frontiers in fundamental science. %While there has been significant progress in the past decades, THz components that allow for an effective control of THz functionalities such as the precise polarization state are still scarce. From this end, 
Spintronic terahertz emitters offer conceptual advantages since the spin orientation in the magnetic layer can be easily controlled either by the externally applied magnetic field or by the internal magnetic field distribution determined by the specific shape of the magnetic elements. Here, we report a switchable terahertz source based on micropatterned magnetic heterostructures driven by femtosecond laser pulses. We show that the precise tunability of the polarization state is facilitated by the underlying magnetization texture of the magnetic layer that is dictated by the shape of the microstructure. These results also reveal the underlying physical mechanisms of a nonuniform magnetization state on the generation of ultrafast spin currents in the magnetic heterostructures. Our findings indicate that the emission of the linearly polarized THz waves can be {switched on and off by } saturating the sample using a biasing magnetic field, opening fascinating perspectives for integrated on-chip THz devices with wide-ranging potential applications.}

\end{abstract}

\maketitle

%\section{Introduction} %to THz radiation (spintronic THz emitter; advantages and promising directions)}

{Terahertz (THz) devices, which operate in the frequency range from 0.1 THz to 30 THz in the electromagnetic spectrum, have significantly impacted various fields such as spectroscopy, sensing, and imaging \cite{SpecAndImg2013}. Many applications in modern life, such as medical diagnosis, security, non-destructive industrial inspection, and non-ionizing imaging, are based on THz radiation. Functional THz components could further advance these technologies. While there has been significant progress in the past decades, THz components that allow for an effective control of THz functionalities such as the {precise polarization state are still scarce} \cite{Khusyainov2021,Yang_AOM2016}. The manipulation of the polarization of THz radiation is a crucial aspect in creating versatile integrated THz devices and circuits. However, the development of such components is hampered by the lack of the required functionalities traditional THz sources can provide: {typically, high-intensity THz sources capable of manipulating the polarization are narrow-band}, limiting possible applications \cite{Kong_2019}. These traditional emitters exploit the electron's charge degree of freedom and rely either on semiconductor based photoconductive antennas or nonlinear crystals \cite{JAP_wu_tutorial}.}
\begin{figure}[b]
    \centering
    \includegraphics[width=1\columnwidth]{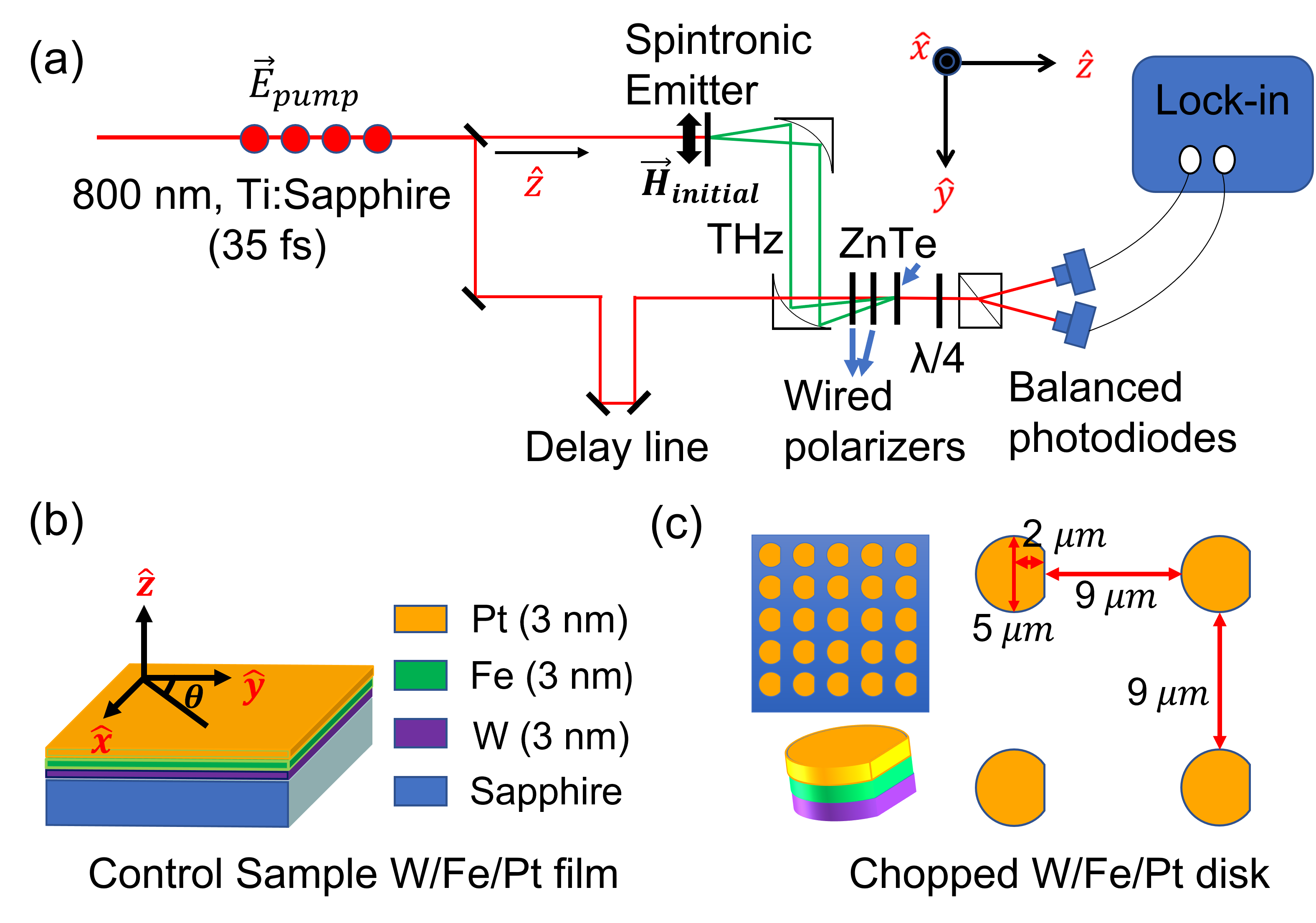}
    \caption{(a) Schematic of a time-domain THz spectroscopy (TDTS) system using electro-optic sampling with a ZnTe crystal and a 800 nm Ti:Sapphire pulse laser with 35 fs pulse duration at a repetition rate of 2 kHz. The time traces of the THz signal are recorded using a delay stage. In our measurements, the laser pulses are polarized in $x$-direction {(out-of-plane)}, while the spintronic emitter is magnetized in $y$-direction and the sample is rotated in the $x/y$-plane [the angle $\theta$ describes the angle between the fixed magnetizing field direction and the rotated sample, see Fig.~\ref{fig1}(b)]. (b) Illustration of sample stack. (c) Dimensions of micropatterned magnetic heterostructures.} 
    \label{fig1}
\end{figure}

{However, %it was realized that adding
the additional spin-degree of freedom of electrons enables functionalities that could further advance THz technologies, e.g., biomedicine, materials science, physics, and chemistry. These \textit{spintronic} THz emitters are based on magnetic heterostructures that comprise ferromagnetic metal/heavy metal layers \cite{JAP_wu_tutorial,Kampfrath_rev_2021}. Upon excitation with a femtosecond laser pulse they emit THz radiation that is both highly efficient and broadband \cite{Kampfrath2013,Kampfrath_Nat2016,Yang_AOM2016,Huisman_Nat2016,Seifert_2017,Jungfleisch_2018,Walowski_JAP2016,Zhou_PRL2018,Torosyan,Nenno_2019,JAP_wu_stripe,Sharma_2021}, which enables the realization of functional THz components with emergent properties based on the spin orientation in the ferromagnetic layer. The THz generation process in spintronic emitters {\cite{JAP_wu_tutorial}} can be understood as a laser-driven ultrafast demagnetization of the ferromagnetic layer \cite{Bigot_PRL1996}, leading to the formation of a spin-polarized current that diffuses in the adjacent heavy metal layer \cite{Battiato2012,Rudolf:2012ky,Nenno2018}, where it is converted into a charge current due to the inverse spin Hall effect \cite{DyakonovJETP1971,Hirsch_PRL1999,HoffmannIEEETM2013}: %$\vec{J} _c \propto \gamma \vec{J}_s \times \vec{\sigma}$
\begin{equation}
\vec{J} _c \propto \gamma \vec{J}_s \times \vec{\sigma},
\label{Eq:ISHE}
\end{equation}
where $\vec{\sigma}$ is the spin-polarization vector, $\gamma$ is the spin Hall angle of the heavy metal layer, and $\vec{J} _c$, $\vec{J} _s$ are the charge and spin currents, respectively.                                                                                                                                                                                           {See supplemental material (SM) for a schematic illustration of the inverse spin Hall effect.}
The time-varying charge current burst then emits THz electromagnetic waves, described by Maxwell's equations. An additional advantage of spintronic THz sources is that micro- and nanofabrication of magnetic thin films are readily available, which opens up the possibility to tailor THz properties by design.}

%Despite of their wide usage, the drawbacks such as complexity on engineering THz emission and high cost of these THz sources cannot be easily overcome. This requires the development of new devices and techniques in generating THz radiation with high efficiency and low cost. 
%In recently years, the development in spintronics and ultrafast demagnetization phenomenon have given birth to the new field of ultrafast spintronics. (add ref). While it is found out that the ultrafast demagnetization phenomenon happens to be in the picosecond time scale, which corresponds to THz frequency in frequency domain, the idea of THz spintronic emitter was come out with. (add ref) Unlike traditional THz sources which only rely on the electron charge. Spintronic THz sources employ an additional degree of freedom of the electron spin. 
%In a typical spintronic emitter, the magnetic heterostructure consists of a ferromagnetic layer and a neighboring heavy metal layer. 
{Previously, microstructured stripes made of magnetic heterostructures were discussed as a means to modify the THz emission characteristics \cite{JAP_wu_stripe} and it was shown that thin film magnetic heterostructures exposed to an inhomogeneous magnetic field distribution can be utilized to create elliptical THz waves \cite{Kong_2019}.} %\sout{However, this approach is not always practical as the emitter requires an external magnetic field which may restrict potential applications. Moreover, the exact position of the sample with respect to magnetic field lines is crucially important and thus this approach may lead to unreliable and unwanted results.}} %\textcolor{blue}{Furthermore, microstructured stripes made of magnetic heterostructures were discussed as a means to modify the THz emission characteristics \cite{JAP_wu_stripe}}. 
{Another approach to control the THz radiation in spintronic emitters is to utilize the remanent magnetization texture such as magnetic vortices and antivortices or magnetic skyrmions \cite{Shinjo930, Nanomagnets}.}
%can be obtained as the ground state of certain micropatterned structures 
 %These structures have been proposed in recent years as potential candidates for a variety of spintronic devices and applications \cite{Locatelli2014, PhysRevLett.97.107204}, and we showed that the spin-Hall effect can be used to excite the underlying low-frequency dynamics \cite{doi:10.1063/5.0006557}. Hence, they present a unique opportunity to merge with THz applications.

{Here, we report a broadband magnetic-field controllable THz source based on the magnetization texture facilitated by micropatterned magnetic heterostructures with broken {mirror} %inversion
symmetry.}% \textcolor{blue}{magnetic-field controlled emission.} %between linearly and elliptically polarized terahertz radiation 
%driven by femtosecond laser pulses. }
{ For this purpose, we use lithographically-defined arrays of a W/Fe/Pt trilayer grown on sapphire. We show that the precise tunability of the polarization state is facilitated by the underlying magnetization texture of the Fe layer that is determined by the exact shape of the elements {and the direction of the applied field with respect to the microstructure}. This interpretation is further supported by micromagnetic simulations. Our results further reveal how a nonuniform magnetization state affects the generation process of ultrafast spin currents in the microstructure and hence the properties of the emitted THz waves. Our findings indicate that {a magnetic switch of linearly polarized THz source can be realized in microstructured geometries}
%chirality as well as ellipticity of the elliptically polarized radiation can be manipulated 
opening fascinating perspectives for integrated on-chip THz devices.}

%\section{Experimental methods}

In the following, we describe the experimental methods including time-domain THz spectroscopy, sample fabrication, and micromagnetic modeling.

%\subsection{Time-domain THz spectroscopy measurements}
{The experimental THz emission measurements are conducted using a standard time-domain THz spectroscopy (TDTS) system using electro-optical sampling with a 300 $\mu $m-thick (110) ZnTe crystal, see Fig.~\ref{fig1}(a). The laser system is a 800 nm Ti:Sapphire regenerative amplifier with a pulse width of 35 fs at a 2 kHz repetition rate and incident laser fluence below $3.5~$mJ/cm$^2$. The THz traces are measured with a time resolution of 50 fs while the pumping laser beam is aligned perpendicular to the sample plane and faces the substrate material, see Fig.~\ref{fig1}.} {An in-plane magnetic field (in $y$-direction) is used to magnetize the sample to initially align the Fe magnetization $M$ (perpendicular or parallel to the direction of symmetry as discussed below).}

\begin{figure}[b]
    \centering
    \includegraphics[width=0.75\columnwidth]{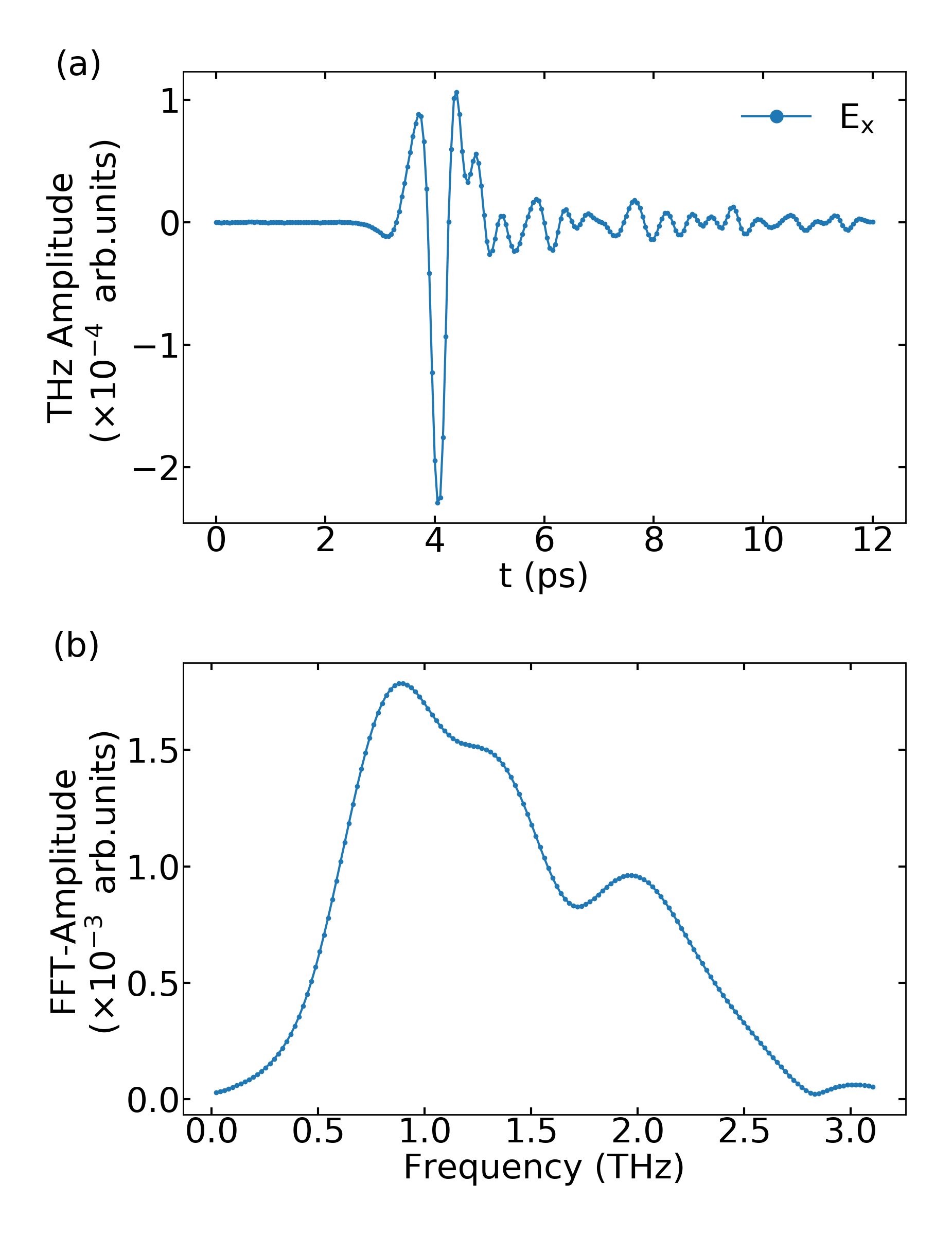}
    \caption{(a) Time trace of the THz emission from an extended W/Fe/Pt trilayer control film. (b) Corresponding intensity of the Fourier-transformed THz signal shown in (a).}
    \label{fig2}
\end{figure}

\begin{figure*}[t]
    \centering
    \includegraphics[width=1.4\columnwidth]{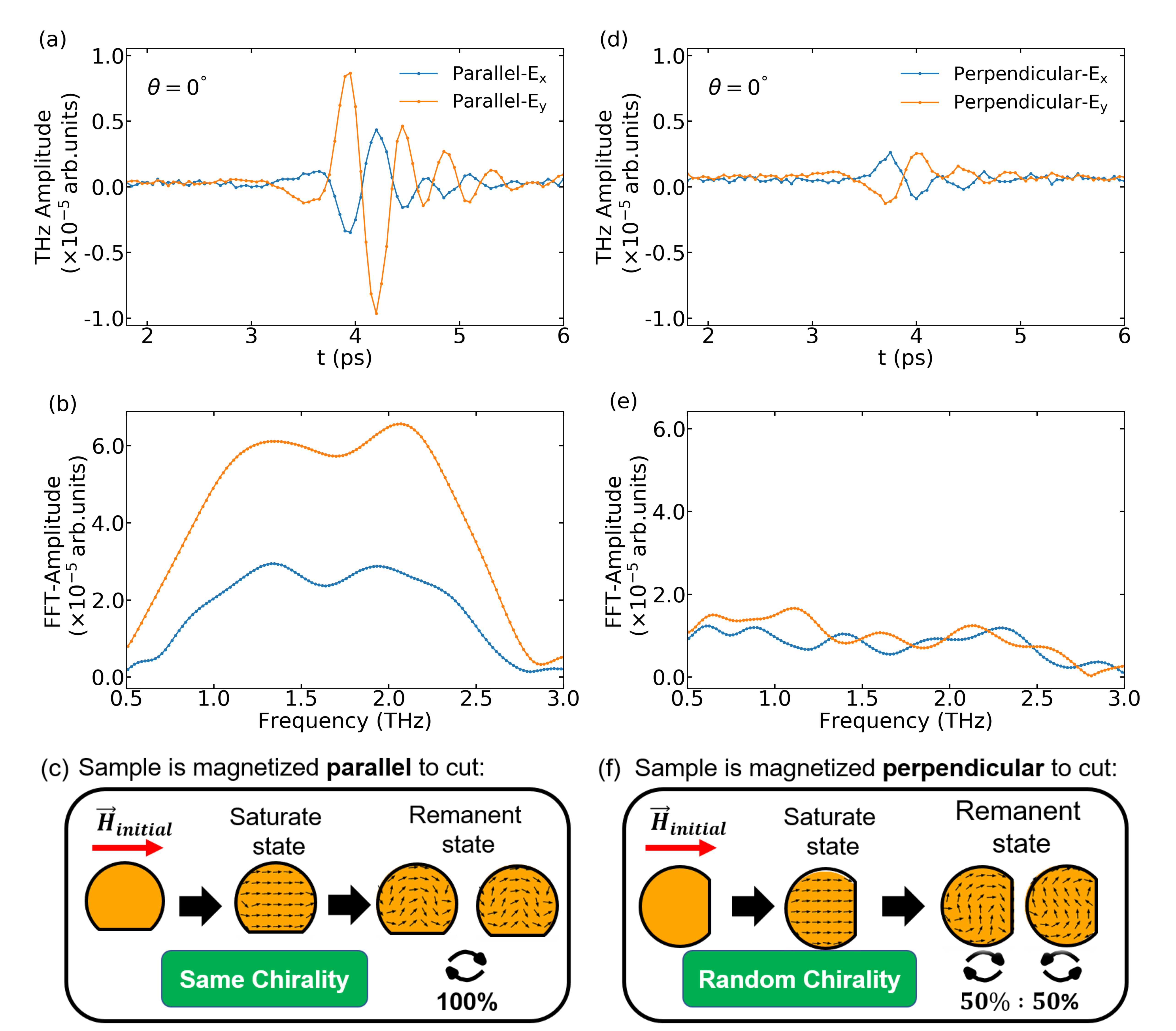}    
    \caption{{Demonstration of {switchable} %\sout{switchability between elliptically polarized and randomly polarized} 
    THz waves by initializing the sample along the direction of {mirror} %inversion
    symmetry (perpendicular to the cut) and orthogonal to the direction of symmetry (parallel to  the cut). For initialization, the angle between sample and magnetizing field (applied in $y$-direction) was $\theta=0^\circ$. (a,d) Time-domain traces of THz emission for the different configurations. (b,e) Corresponding Fourier-transformed spectra. (c,f) Schematic of the magnetization texture in remanent state when the initializing field is applied parallel or perpendicular to the cut. Ideally, one would expect that all disks have the same chirality when initialized parallel to the cut (c), whereas a random distribution of chirality would be expected when initialized perpendicular to the cut. For micromagnetic modeling results we refer to the SM.}}
    \label{fig3}
\end{figure*}

%\subsection{Sample fabrication}
{To increase the intensity of the emitted THz radiation we sandwich the ferromagnetic layer (Fe) between two heavy metals with opposite spin Hall angles (W and Pt), see Fig.~\ref{fig1}(b). Since the thickness of the used materials is small (nanometers), the THz radiation originating from the two heavy metal layers will have negligible phase difference. {The samples were fabricated in the following fashion: we use optical lithography by a laser writer (Heidelberg MLA 100, minimum feature size: 1 $\mu$m) to pattern an array of chopped disks (diameter: 5 $\mu$m, asymmetric cut of the disk from its center: 2 $\mu$m) with a separation of 9 $\mu$m between inner edges on top of sapphire substrate.} 
%\sout{The samples were fabricated in the following fashion: Optical lithography was used to pattern an array of chopped disks (diameter: 5~$\mu$m, asymmetric cut of the disk from its center: 2~$\mu$m) with a separation of 9~$\mu$m from center to center on top of sapphire substrate.}} 
{The sapphire substrate is $5~\mathrm{mm} \times 5~\mathrm{mm}$ with 0.5 mm thickness in the C-plane (0001).}
%{The cut of the disk is aligned along one side of the substrate to be able to easier initialize the direction of the initially applied magnetic field with respect to the cut.}
{The cut of the disks is patterned parallel to one side of the substrate to easily align the external magnetic field with respect to the microstructured disk array. The detailed configuration is shown in Fig.~\ref{fig1}(c). The metal layers W (thickness: 3 nm), Fe (thickness: 3 nm), and Pt (thickness: 3 nm) are grown using dc magnetron sputtering in an Ar atmosphere at 3 mTorr and 20 sccm gas flow (base pressure $ < 10^{-7}$ Torr) on double-side-polished sapphire substrates, followed by a lift-off process. 
{Microscope and atomic force microscopy images of the measured samples are shown in the SM.}} A control W/Fe/Pt trilayer film was grown at the same time in the same deposition step.

%\textcolor{red}{Weipeng, can you please fill in the fabrication details? How were the samples grown, what are growth parameters, materials, thickness, dimensions, etc.?}

%\subsection{Micromagnetic simulations}
{We perform micromagnetic simulations using the graphic processing unit (GPU) accelerated software mumax3 \cite{mumax3} to determine the magnetization state of the microstructured disks. We simulated a primitive cell of $2\times 2$ lattice sites with periodic boundary conditions in the plane, and considering 3 repeated cells along each direction for the initial calculation of the dipolar field.}
%{which were repeated to 3 replicas along both axes.}
{Each disk with 5 $\mu$m diameter and 3 nm thickness is cut 500 nm from one side to replicate the experimental conditions. The lattice is discretized in square-based elemental cells of $5\times 5 \times 3$~nm$^3$ and the whole structure consists of $2400\times 2400\times 1$ cells. \footnote{The cell size is chosen larger than the exchange length of iron ($\approx$2.4 nm) to reduce the computation time because our structure is fairly large.} %(Not sure if you want to explain this part: The cell size is larger than the exchange length of iron (2.4 nm). This is set to reduce the simulation time as the whole lattice is rather large). 
The magnetic parameters used for Fe are as follows: saturation magnetization $M_\mathrm{S}= 1710$~kA/m and exchange stiffness $A_\mathrm{ex}=21.2 \times 10^{-12}$~J/m. % and Gilbert damping parameter $\alpha= 0.0013$. 
}
%\sout{In order to initialize the magnetization in the sample, an initial magnetic field was applied.}
{ For each applied magnetic field direction (parallel or perpendicular to the cut, respectively) we started with disks with random magnetization in each cell that were exposed to an external magnetic field with a magnitude of 8 mT, and then removed the field to obtain the final magnetization equilibrium state of the lattice. {However, a field of 50 mT was used to initialize the samples in the experiments and all measurements were done in the remanent state.} } %\sout{For this purpose, we used the minimize and then relax methods in mumax3 after each field step and finally we let the simulations evolve for 20 ns.}

$$
%\vec{J} _c \propto \gamma \vec{J}_s \times \vec{\sigma}
$$
Figure~\ref{fig2}(a) shows the time-domain trace of the THz signal (in $x$-direction) generated by the W/Fe/Pt trilayer control film. In-plane angular dependent measurements agree with a spin-to-charge current conversion process by means of the {inverse spin Hall effect, see SM.}
%{ Moreover, we tested the dependence of the helicity of the pump light on the THz emission (not shown here). However, no dependence was observed}
{Moreover, we observe that the THz emission is independent from the helicity of the pump light, shown in Fig.~S3 in the SM}
{ implying that inverse spin orbit torque \cite{Nemec_NatPhys_2012}, inverse Faraday effect \cite{Kimel_Nat_2005}, or inverse Rashba Edelstein effect \cite{Huisman_Nat2016,Jungfleisch_2018} can be ruled out as possible sources of the observed THz emission} {\footnote{Please note that when the sample is firstly magnetized and the field is then removed before the TDTS measurement, we also observe a contribution of the THz signal in $y$-direction which likely stems from a not perfectly magnetized sample in $y$-direction and/or an applied magnetizing field that is not large enough to saturate the Fe layer}. 
The corresponding Fourier spectrum of this reference signal [shown in Fig.~\ref{fig2}(b)] demonstrates that a broadband THz beam is generated.}

{In the following, we discuss the main observation of our work -- the  manipulation of the ultrafast spin current generation driven by femtosecond laser pulses facilitated by the underlying magnetization texture in the micropatterned magnetic heterostructure. As will be shown below, this control is robust and is enabled solely by the effective magnetic field distribution of the microstructure. Once the magnetic texture is initialized by an external biasing field,}
%\sout{ with respect to the characteristic lateral symmetry-breaking feature,}
{ no additional external magnetic field is required; this is particularly useful for applications since the THz device itself does not require to be powered.}

{Our results show that it is possible to switch on or off %\sout{between elliptically} 
linearly polarized %\sout{and randomly polarized } 
THz waves by initializing the sample in either of two orthogonal directions, in the following called ``parallel'' and ``perpendicular'' to the cut. {The peak amplitude of the THz signal of the chopped disks samples is about 1/20 of that of Fe/Pt bilayer film of the same thickness. This ratio is of the same order as the ratio of the two-dimensional hybrid metal halides with respect to the control sample ($E_\mathrm{NiFe/Halides}/E_\mathrm{NiFe/Pt}\approx$1/15) \cite{Cong2021}. Two-dimensional hybrid metal halides are chiral magnetic materials for the generation of THz waves with asymmetric magnetic field dependence.} For the initialization process a constant magnetic field is applied in the two directions, subsequently removed, and the measurements are performed without a magnetic field being applied. Since we expect a curled magnetization texture with magnetic moments lying in any direction in the $x/y$-plane, we record the THz electric field amplitude in $x$-direction, $E_\mathrm{x}$ and in $y$-direction, $E_\mathrm{y}$. Figure~\ref{fig3} shows the THz signals in the time domain and the corresponding fast-Fourier transform (FFT) spectra in the frequency domain in $x$- and $y$-direction when the sample is magnetized parallel and perpendicular to the cut, respectively. As is apparent from the figure, a negligibly small THz signal is observed when the sample is initially magnetized perpendicular to the cut. On the other hand, a clear signal is observed in both $x$- \textit{and} $y$-direction when the sample is magnetized parallel to the cut. The ability to switch between the two states is reproducible and robust evidencing that this effect must originate from the magnetization configuration in the microstructured disks with broken {mirror} %inversion 
symmetry.}

%\textcolor{red}{Can we say something about the phase between Ex and Ey at this point? --> inverted signal?}

\begin{figure}[t]
    \centering
    \includegraphics[width=0.8\columnwidth]{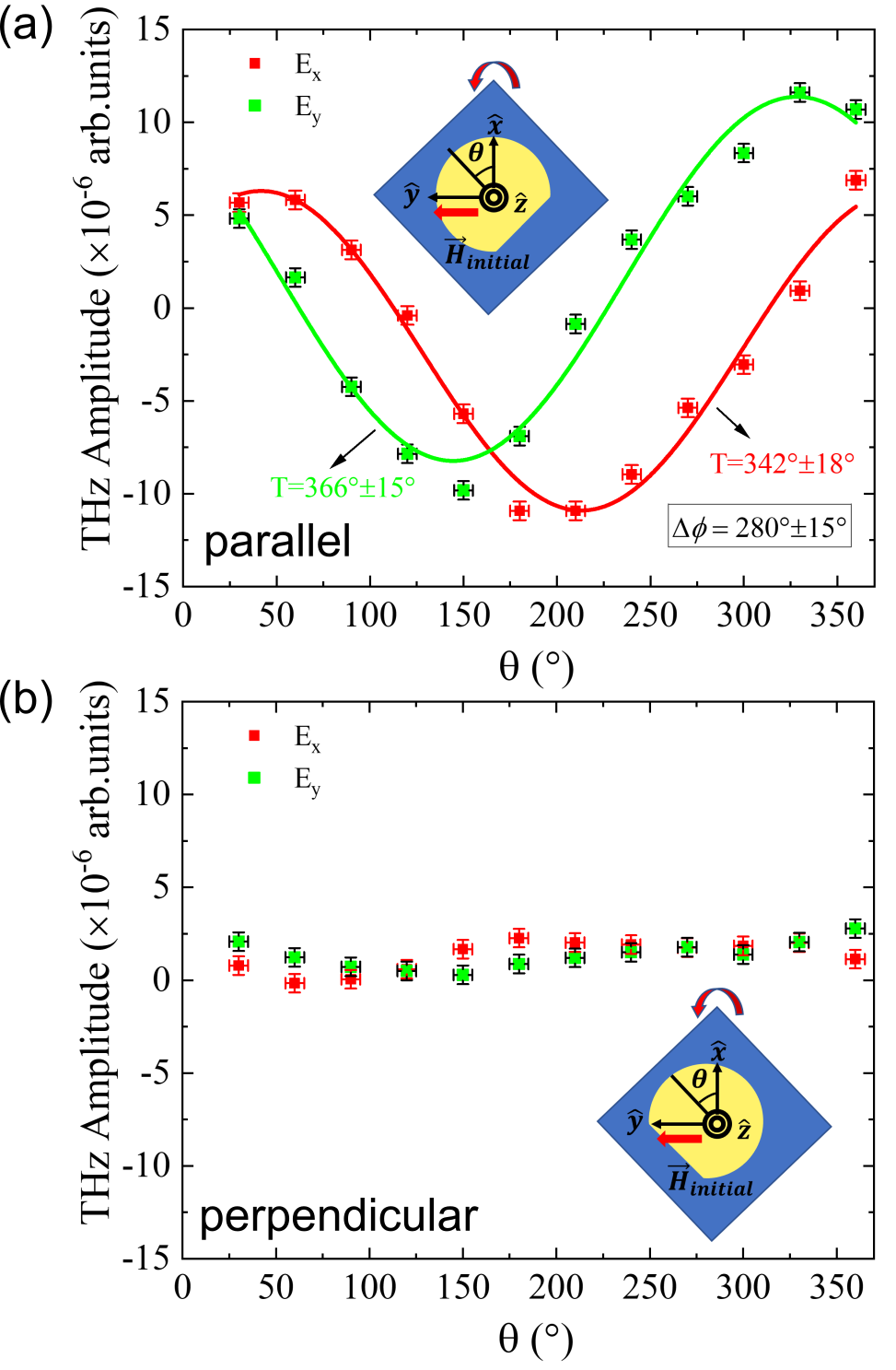}
    \caption{(a) In-plane angular dependence of the THz amplitude generated from micropatterned chopped disks initially magnetized parallel to the symmetry-breaking cut. $E_\mathrm{x}$ becomes maximum/minimum when $E_\mathrm{y}$ is minimum/maximum. The solid lines show a sinusoidal fit to the experimental data (symbols). The inset shows a schematic of disks, the initially applied magnetic field direction, and the used coordinate system. (b) Corresponding results for the case when the chopped disks are initially magnetized perpendicular to the symmetry-breaking cut. The THz amplitude was chosen at the same delay time for different angles. A negligibly small THz signal was detected; no significant angular dependence was observed for the two orthogonal components $E_\mathrm{_x}$ and $E_\mathrm{y}$ in this configuration.}
    \label{fig4}
\end{figure}

{To further support our interpretation and to provide guidance on the exact underlying magnetization texture of the disks, we use micromagnetic modeling based on mumax3 (SM) \cite{mumax3}. A schematic illustration explaining our observations is shown in the lower panel of Fig.~\ref{fig3}; actual simulation results can be found in the SM. When the sample is firstly magnetized parallel to the cut, due to the broken {mirror} %inversion 
symmetry, only one of two possible magnetization configurations (either clockwise or counterclockwise depending on whether the initial field is applied in positive or negative $y$-direction) predominantly exists in the remanent state, see schematic in Fig.~\ref{fig3}(c). In other words, all magnetic microelements in the array feature the same chirality with magnetic moments identically distributed in the $x/y$-plane (please note that the laser spot size is sufficiently large to excite a multitude of disks). This in turn leads to a THz emission with {fixed linear} %\sout{certain} \sout{elliptical} 
polarization, and we can observe both $E_\mathrm{x}$ and $E_\mathrm{y}$ components, see Fig.~\ref{fig3}(a). However, the situation drastically changes when the sample is magnetized perpendicular to the cut -- the direction of {mirror} %inversion
symmetry of the disks: now two configurations emerge in the remanent state with equal probabilities, }{resulting in a random distribution of chirality in each of the disks in the array,}{ see schematic shown in Fig.~\ref{fig3}(f). If both chiralities coexist in the sample, an overall cancellation of THz emission due to %\sout{opposing effects} 
{the near-zero net magnetization} occurs. This is the reason why we observe a negligibly small signal when the sample is initialized perpendicular to the cut, see Figs.~\ref{fig3}(d,e). Please note that we cannot rule out patterning imperfections in the actual samples, which may lead to a deviation from the theoretically expected 100\% or 50\% of the }%\sout{same chirality in every disk} 
{circularity probability distribution in the disks}%\textcolor{red}{I think we can keep what it was}
{ when initialized parallel or perpendicular to the cut, respectively.}

{To further confirm our interpretation, we  measure the dependence of the THz emission on the in-plane sample rotation angle $\theta$ [see Fig.~\ref{fig1}(b)] around the $z$-axis for  initial magnetizing directions parallel and perpendicular to the cut, see Figs.~\ref{fig4}(a,b). %\sout{The corresponding results when the sample is magnetized perpendicular to the cut are omitted here as no THz emission was detected in that case. For more information we refer to the SM.}
Figures~\ref{fig4}(a,b) show the amplitude of the two orthogonal components of THz radiation $\mathrm{E_x}$ and $\mathrm{E_y}$ taken at identical delay times when the signal peaks in the time domain. %\textcolor{blue}{for the two different magnetizing directions}. 
As we would expect, the $x$-component is maximum when the $y$-component is minimum, and vice versa. The experimental data in Fig.~\ref{fig4}(a) (symbols) are fitted with sinusoidal functions (solid lines) that confirm the periodic behavior: fits to the two components give a  periodicity that is close to 360$\mathrm{^\circ}${-- a characteristic of ferromagnetic spintronic emitters relying on the inverse spin Hall effect \cite{JAP_wu_tutorial}.}  {{As outlined in the detailed analysis presented in the SM, there may be an elliptical contribution to the THz signal. However, this effect is small.}} {In contrast to the angular dependence when the disks are magnetized along the cut, negligible THz signal and angular variation are detected when the disks are initialized perpendicular to the cut, Fig.~\ref{fig4}(b).}}  %\sout{a period of approximately} {approximately one period oscillation in} {360$^\circ$ implying that the magnetic moments have an orientation in the $x/y$-plane.}

{Finally, we note that an additional tuning variable could be a spatial variation of the thickness of the layers or their conductivities that would enable the control of the exact phase difference as well as the relative strength of the two orthogonal components.} %{\textcolor{blue}{As is seen in the detailed analysis of the polarization of the THz radiation, the emitted THz wave turns out to be slightly elliptically polarized with a very small ellipticity. Here we still refer to as linear polarization. We however enphasis that}} 
{This approach may lead to a reliable generation of elliptically polarized THz waves with tunable ellipticity and polarization direction from spintronic emitters \cite{Kong_2019}.}

{We demonstrated an effective way to %\sout{manipulate the broadband THz emission properties including chirality, ellipticity, and amplitude exploiting conceptual advantages }
{control and switch the linearly polarized THz radiation} from femtosecond laser pulse driven spintronic microstructure emitters.}
%\sout{ The precise tunability of the polarization state is facilitated by the underlying magnetization texture of the magnetic layer that is dictated by the effective internal magnetic field distributing determined by the shape of the microstruture.}
{We showed that the magnetic texture in the ferromagnetic layer controlled with the shape of the microstructure dictates the emitted THz waves.} {Our findings indicate that %\sout{ chirality as well as ellipticity of the elliptically} 
the emission of a polarized THz radiation can be controlled without an external magnetic field applied during the THz generation process with carefully designed microstructures. This goes beyond a recent proposal} {to harness permanently applied inhomogeneous  magnetic fields \cite{Kong_2019}, opening fascinating perspectives for integrated on-chip THz devices with wide-ranging potential applications. Our results also reveal the underlying physical mechanisms of a nonuniform magnetization state on the generation process of ultrafast spin currents and the subsequent conversion in THz charge transients by the inverse spin Hall effect in the magnetic heterostructures.} 
{The proposed approach could potentially be used in spintronic emitters for the generation of elliptically polarized THz waves with tunable polarization, chirality, and ellipticity.}
%textcolor{blue}{which implies potential application of microstructures in the generation of elliptically polarized THz waves with tunable polarization, chirality and ellipticity.}}

See the supplementary material for details on the inverse spin Hall effect, results of micromagnetic simulations, helicity-dependent THz emission, angular dependence of the control sample, microscope and atomic force microscopy images of the samples under investigation, and angular dependence of the total THz electric field.

\begin{acknowledgments}
This work was supported by the National Science Foundation under Grant No. 1833000 and the University of Delaware Research Foundation (UDRF). Additional support received from the National Science Foundation through the University of Delaware Materials Research Science and Engineering Center, DMR-2011824. 
Work performed at the Center for Nanoscale Materials, a U.S. Department of Energy Office of Science User Facility, was supported by the U.S. DOE, Office of Basic Energy Sciences, under Contract No. DE-AC02-06CH11357.
%Use of the Center for Nanoscale Materials, an Office of Science user facility, was supported by the U.S. Department of Energy, Office of Science, Office of Basic Energy Sciences, under Contract No. DE-AC02-06CH11357. 
H. W. %\sout{also} 
acknowledges support from the Department of Energy, Office of Science, Office of Basic Energy Sciences, %\sout{under contract no. DE-SC0012509}
{Materials Sciences and Engineering Division.}.
\end{acknowledgments}

The authors have no conflicts to disclose.

%\nocite{*}
\bibliography{aipsamp}% Produces the bibliography via BibTeX.

\end{document}